# WEEP: A method for spatial interpretation of weakly supervised CNN models in computational pathology

Abhinav Sharma[1], Bojing Liu[1], Mattias Rantalainen[1*]

[1] Department of Medical Epidemiology and Biostatistics, Karolinska Institutet, 17165, Sweden

* corresponding author: mattias.rantalainen@ki.se

# Abstract

## Background

Deep learning enables the modelling of high-resolution histopathology whole-slide images (WSI). Weakly supervised learning of tile-level data is typically applied for tasks where labels only exist on the patient or WSI level (e.g. patient outcomes or histological grading). In this context, there is a need for improved spatial interpretability of predictions from such models.

## Results

We propose a novel method, Wsi rEgion sElection aPproach (WEEP), for model interpretation. It provides a principled yet straightforward way to establish the spatial area of WSI required for assigning a particular prediction label. We demonstrate WEEP on a binary classification task in the area of breast cancer computational pathology.

## Conclusion

WEEP is easy to implement, is directly connected to the model-based decision process, and offers information relevant to both research and diagnostic applications.

# Background

Deep learning-based models have demonstrated high performance in a range of prediction tasks in the digital pathology domain. Due to hardware constraints, a common strategy is to divide the gigapixel-size whole slide images (WSIs) into smaller patches (i.e. tiles), which are subsequently modelled by e.g. deep Convolutional Neural Network (CNN) models (1–3). In situations where labels are only available on the WSI level, either due to the absence of pixel-level annotations or the presence of labels exclusively on the WSI level (e.g. patient outcomes, time-to-event, treatment response), weakly supervised learning is often applied in the domain of computational pathology (4). This type of precision pathology applications (5) are of high value in medicine, as artificial intelligence has the potential to go beyond what a human expert can see in a histopathology specimen and thus extract and provide new or additional information with high clinical relevance.

Weakly supervised learning has demonstrated broad success in precision pathology and computational pathology in general, also in the presence of label noise. However, the interpretation of these models is not straightforward, despite the importance of the interpretation in both research applications and clinical decision support applications. Conventional tools for the interpretation of deep CNN models, such as class activation maps (CAM) (6) and Gradient-weighted class activation maps (Grad-CAM) (7), are focused on pixel-level local interpretability (within tile), which do not have a direct interpretation with respect to the predicted label at the WSI level (across all tiles). Recently, trainable attention-based neural networks (8) have been introduced in histopathology image analysis (9). However, the attention weights lack a direct relationship to the decision problem and can therefore not directly be used for determining the set of tiles (i.e. area) required for a particular classification label.

In precision pathology, spatial interpretability has the potential to highlight areas (at tile level) with the most relevant tissue morphology for the particular modelling task at hand. Especially in the context of weakly supervised CNN models of tile-level data, methodology for spatial interpretability linked to the prediction problem has not been available.

To address this problem, we propose a novel methodology, the Wsi rEgion sElection aPproach (WEEP), enabling ascertainment of WSI regions that are required for a positive classification label at the slide-level, in the common scenario of tile-based weakly supervised learning. WEEP output has a direct connection to the prediction problem and is directly driven by the model and the data. The approach is easy to implement and has a direct interpretation with respect to the prediction problem and is therefore relevant in both research and clinical application contexts.

To illustrate the method, we apply WEEP to a binary classification task, the classification of histological grade 1 and 3 in breast cancer. Histological grade is a well-established histomorphology-based prognostic marker used to determine the prognosis and treatment planning for the breast cancer patients (10). Deep learning based binary classification of histological grade using MIL-based pooling for tile-to-slide level aggregation has been demonstrated in several studies previously (11–14). WEEP can also contribute to understanding the WSI areas and morphological patterns for novel histo-morphological biomarker discovery. Similarly, the method can be further extended to other precision pathology tasks that include survival prediction, treatment prediction among others.

In this study, we describe the WEEP methodology and apply it to identify classification-essential WSI regions from three different tile-to-slide level aggregation functions: a trivial mapping function based on the 75th percentile of tile-level prediction scores from CNN models, a tile-level attention scores from a trainable attention-based model, and a transformer-based aggregation model..

# Methods

## WEEP method description

Multiple Instance learning (MIL) is one of the standard frameworks for weakly supervised learning problems in computational pathology (15). In a standard MIL framework, a bag (represented by the WSI) contains instances (represented by the tiles), and if at least one of the instances is classified as positive then the whole bag is classified as positive. However, in real-world computational pathology applications, the bag level label is typically assigned based on a function that accounts for more than a single tile in order to achieve optimal prediction performance. The MIL framework can be divided into two steps: first, the tile-level feature representations/prediction scores are obtained from a base CNN model and second: these feature representations/prediction scores are aggregated to provide the WSI prediction score using a tile-to-slide aggregator function or model. Different tile-to-slide level aggregator functions have been used in different weakly supervised classification scenarios (16).

WEEP exploits a fundamental property of MIL models applied to tile-level instances of the histopathology WSIs. Specifically, we assume the existence of a model that provides tile-level predictions, that allow the ranking of tiles with respect to the prediction task, based on e.g. predicted class probabilities, such that two instances $x_i$ and $x_j$, $p(\text{class}=1 \mid x_i) > p(\text{class}=1 \mid x_j)$ implies $x_i$ is more likely than $x_j$ to belong to class 1, conditional upon the model and the tile. In the case of a simple tile-to-slide mapping function, e.g. mean, median or 75-percentile, of the tile distribution, we can simply use the tile class probability prediction. However, in cases where we have a second trainable model for slide-level predictions that has attention weights, these weights can also be considered as the ranking metric.

Irrespective of the ranking metric, a backward selection approach can be applied, which provides guarantees that the selected subset is the maximum set of tiles needed for positive classification on the WSI level. Backward selection is a common strategy that has been used for variable selection in predictive modelling tasks (17). Here, we apply it to identify the set of instances (tiles) required for a positive classification label (Table 1). In our empirical evaluation results, we consider tile-level predicted class probability as well as attention weights as ranking metrics. The WEEP algorithm allows us to determine the set of tiles directly required to assign the WSI label (Figure 1).

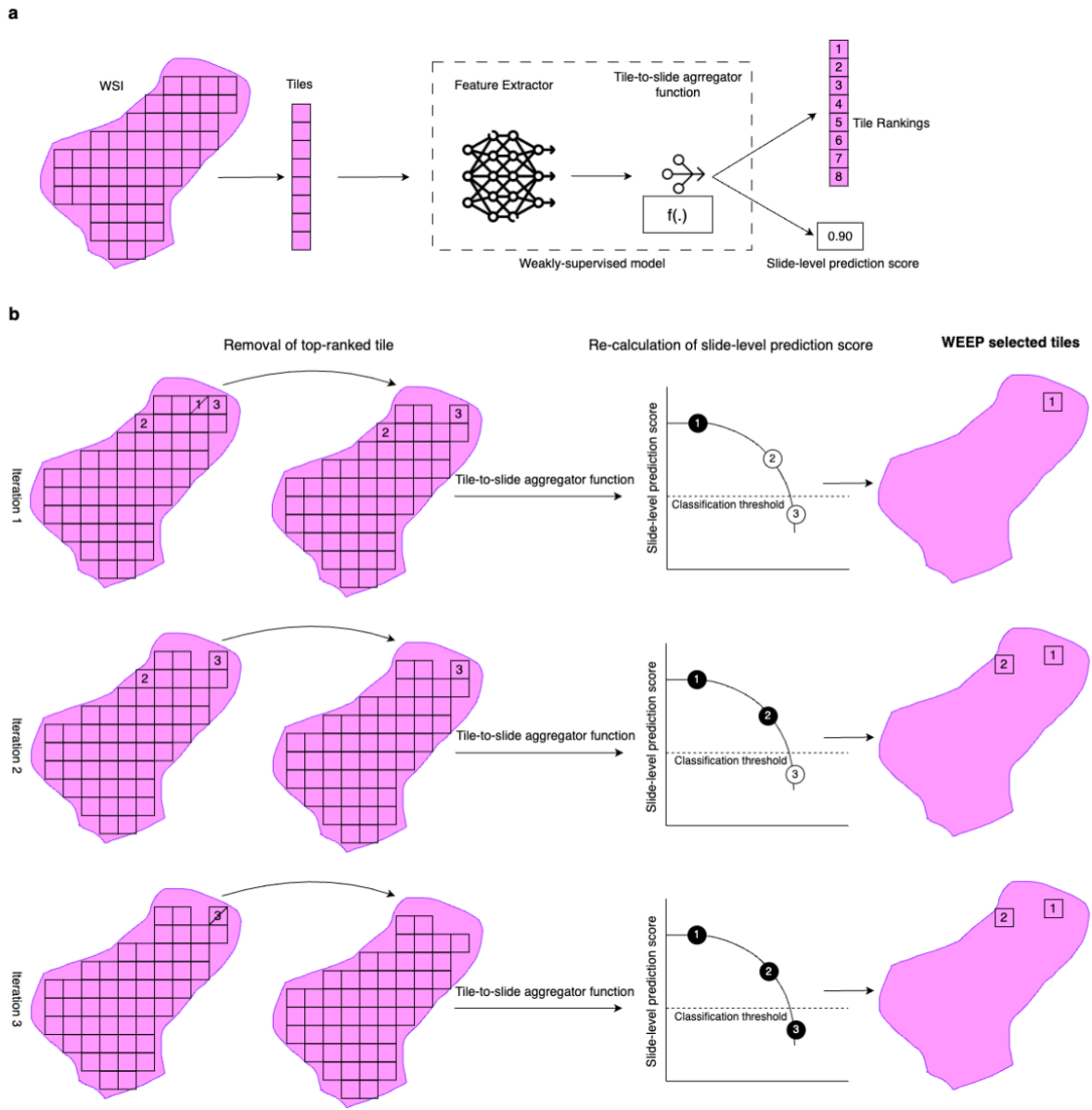

Figure 1: Overview of the WEEP methodology. a Demonstrating a weakly-supervised learning scenario for histopathology image analysis. Firstly, whole-slide image (WSI) is divided into small patches called tiles. Then, tiles are provided as input to the weakly-supervised model that is majorly composed of two modules i.e. Feature Extractor and Tile-to-slide aggregator function. The feature extractor module is used to extract low-dimensional representation features from each tile and then tile-to-slide aggregator function (either based on summary statistics or a separate trainable model) is used to provide slide-level prediction score. The tile-to-slide aggregator function utilises the ranking of the tiles based on tile-level prediction scores

or optimised attention scores from trainable attention module. b Demonstration of the WEEP methodology that applies stepwise backward selection of the tiles based on the ranking retrieved from the tile-to-slide aggregator function. In the first iteration, highest ranked tile is removed from the WSI and then tile-to-slide aggregator function is applied on the remaining tiles to retrieve the slide-level prediction score. The following iterations are followed until the slide-level prediction score reaches below the classification threshold (iteration 3 in the example). The removed tiles are the selected tiles through WEEP that are associated with the positive classification of the WSI.

```
1. for a model M; and a WSI with a set of x tiles:
2.    x_selected = {}
3.    sort the set of tiles x, based on p(NHG=3 | x_i), or
      attention weights (a_i) in descending order
4.    calculate the slide-level prediction score, P = f(x, M)
5.    while P >= O:
6.       x_selected = x_selected ∪ {x_i}(store the selected tile)
7.       x = x \ x_i (remove the selected tile from the set)
8.       update the slide-level prediction score, P = f(x, M)
9.    x_selected is the set of tiles that constitute the tiles/region
      driving the classification of the WSI
```

**Table 1:** Outline of the steps in the WEEP algorithm. (*O represents the established decision threshold*)

## Study materials

In this study, we included patients from the SöS-BC-4 cohort, collected from the Södersjukhuset (South General Hospital) in Stockholm, Sweden from the year 2012-2018. One Hematoxylin and Eosin (H&E) stained WSI scanned at 40X magnification level was considered from each patient. WSI preprocessing steps were followed as described in (13). We only included tiles predicted as invasive cancer (tile size: 598x598 pixels at 20x resolution) in further analyses.

The models were optimised and validated on the subset of SöS-BC-4 (N = 1695) using 5-fold cross-validation (cv). The dataset was split into CV training and CV test set for each cv fold. Further, the CV training set was split into the training set, and the tuning set(14). Each data split was stratified by the clinical NHG. The SöS-BC-4 data split is shown in Supplementary Figure 1.

Description *of the four weakly-supervised modelling strategies*

*In this study, we included four weakly supervised modelling strategies. In the first modelling strategy, w*e considered the Resnet-18 CNN model architecture (18) as the tile-level classification model. Pretrained model from Imagenet was used to initialise the model weights (19) and was fine-tuned on the feature extractor training set as the binary histological grade 1 vs 3 classification model. The 75th percentile of the tile-level prediction scores from the CNN models was used to provide the slide-level prediction score as the first aggregation function (12).

For the second modelling strategy, the tile-level feature vectors were extracted from the average pooling layer of the fine-tuned Resnet-18 models for tiles in the attention module training set and tuning set. The trainable attention-based MIL (atten-MIL) model was considered as the second tile-to-slide aggregation function. It includes the attention layer that is a trainable layer inspired by (15), which optimises an attention weight to each tile-level feature vector and provides a slide-level prediction score by performing a weighted average of all tile-feature vectors belonging to that slide. The third modelling strategy included extraction of features for the CV training set using the publicly available foundation model called UNI (Chen et al. 2024). The transformer-based MIL model called TransMIL (Shao et al. 2021) was considered as the tile-to-slide aggregator function. The fourth modelling strategy included atten-MIL as the tile-to-slide aggregation function using the features extracted from UNI. Optimisation of the four modelling strategies is described in the Supplementary.

*Validation on the CV test set*

The optimised weakly-supervised models were validated on the CV test set in each CV fold. We further aggregated the slide-level prediction scores from the five CV test sets and evaluated the optimal classification threshold to classify the WSIs as Grade 3 and 1 using the Youden's statistics.

**Quantitative and Visual analysis of selected regions in different modelling strategies**

We considered the assessment of the WSI regions contributing to the classification of the histological grade 3 WSIs (n = 543). We demonstrated the iterative backward selection approach of the WEEP methodology as the line plot (referred to as the *WEEP plot*). Further, we observed the distribution of the percentage of selected tiles for each WSI in the CV test set using the histogram plots. The selected tiles using WEEP for the example WSIs were visualised over the tumour mask on the low-resolution WSI as the binary mask. All the plots were created using the package matplotlib (v.3.6.2) (20) in python (v.3.10.8).

# Results

**Evaluating the selected regions from different modelling strategies**

First, we applied WEEP to analyse two different tile-to-slide aggregation functions for the binary classification problem of NHG 1 vs 3, and visualised the results as the WEEP

plot(Figure 2a, 2c, 2e, 2g). The results reveal the trajectories of the WEEP backward selection and the corresponding WSI-level prediction score, the horizontal line shows the decision boundary established on the full dataset. Next, we investigated the distribution of the proportion of tiles that were selected by WEEP in each WSI (Figure 2b, 2d, 2f, 2h). For the ResNet-18 model with 75th percentile tile-to-slide aggregator function, a mean percentage of 32.34% (95% CI: 29.69% - 34.99%) was observed as the average percentage of tiles contributing to the histological grade 3 classification in each slide. For the ResNet-18 model with the attention module, we observed the mean percentage of selected tiles as 44.97% (95% CI: 41.69% - 48.26%). For the UNI as the feature extractor, we observed the mean percentage of selected tiles as 63.88 (95% CI: 60.70% - 67.07%) and 71.97% (95% CI: 68.66% - 75.29%), with the atten-MIL and the TransMIL model respectively. Further, we have reported the classification performance of the four weakly-supervised modelling examples in the Supplementary Figure 2.

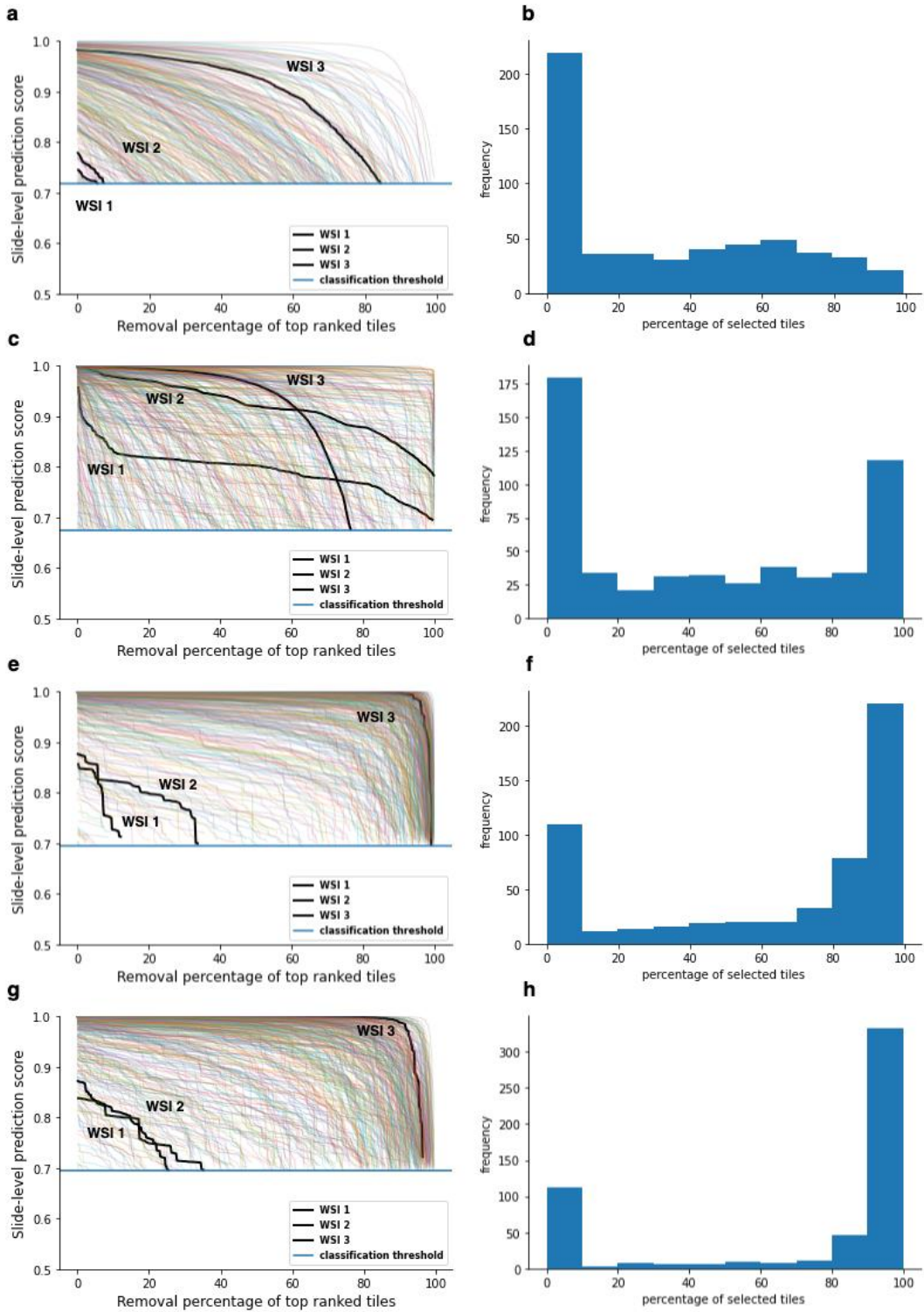

**Figure 2: Application of WEEP to a binary classification problem.** We have randomly selected and highlighted three clinical NHG 3 WSIs with each WSI randomly selected from the ranges <20%, >20% & <80% and >80% of the selected tiles through WEEP using the UNI as feature extractor and TransMIL as the tile-to-slide aggregation model. They are highlighted (black) in plots **a**, **c, e, g**.

**a,c,e,g** WEEP plot showing the change in slide-level prediction score with the step-wise backward removal of top-ranked tiles until the slide-level prediction score reaches the classification threshold. a, ranking of the tiles was based on tile-level prediction scores from the fine-tuned Resnet-18 model and slide level score was determined using the 75th percentile tile-to-slide aggregation function. c, ranking of the tiles was based on the attention scores from the attention-based MIL (atten-MIL) model with fine tuned Resnet-18 as the feature extractor. e, ranking of the tiles was based on the attention scores from the TransMIL model with UNI as the feature extractor and g, ranking was based on attention scores from atten-MIL model with UNI as the feature extractor. **b,d,f,h** Distribution of the percentage of selected tiles for each WSI from WEEP when applied to the: b tile-level prediction scores with 75th percentile tile-to-slide aggregation function, d attention scores from the attention-based MIL (atten-MIL) model with fine tuned Resnet-18 as the feature extractor, f attention scores from the TransMIL model with UNI as the feature extractor, h attention scores from the atten-MIL model with UNI as the feature extractor.

## Visualisation of the selected regions from different modelling strategies

To provide an example of the resulting spatial interpretation, we visualised the selected regions for different modelling strategies (Figure 3). Revealing the spatial localization and WSI regions that are directly linked with the classification of histological grade 3 WSIs.

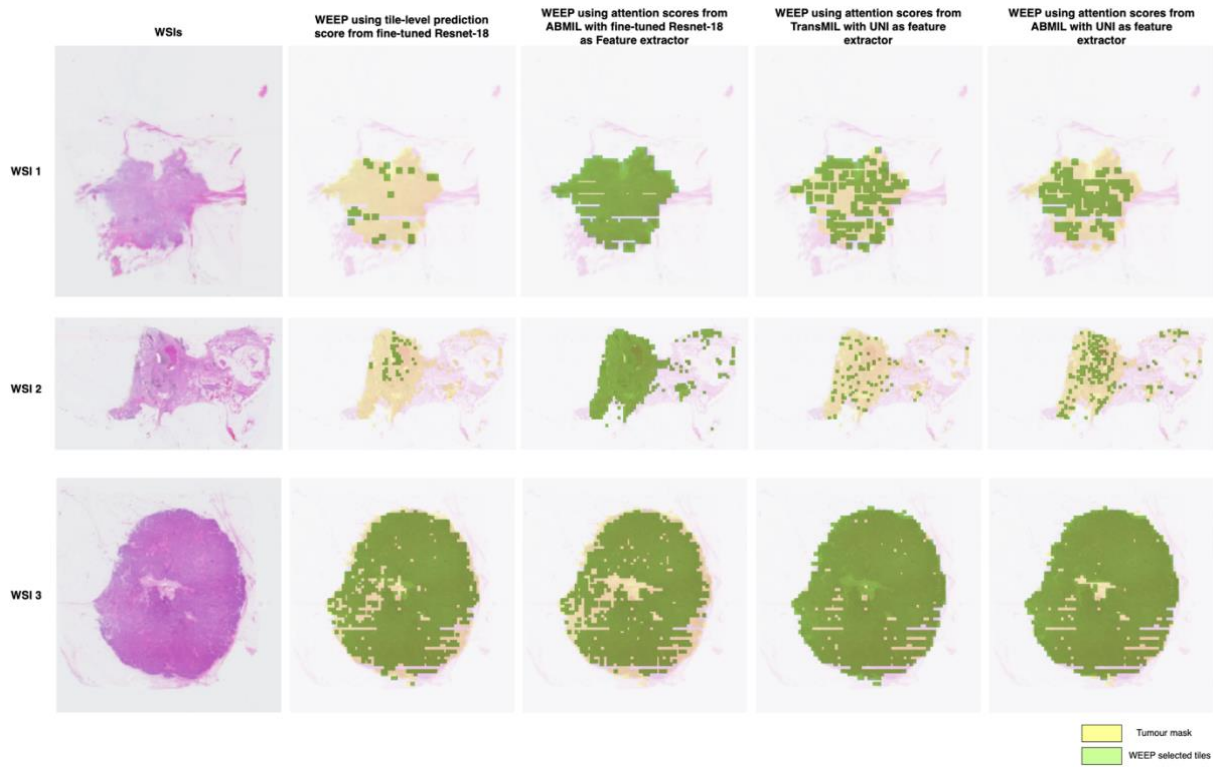

**Figure 3:** Binary masks of the selected tiles through WEEP over the tumour mask on the original WSI. Randomly selected three clinical NHG 3 WSIs from Figure 1 have been demonstrated here. The second column represents the binary masks of the selected tiles by applying WEEP to the tiles ranked by tile-level prediction scores using the fine-tuned Resnet-18 model. The third column represents the binary masks of the selected tiles by applying WEEP to the tiles ranked by attention scores from attention-based multiple instance learning (atten-MIL) model using features extracted from fine-tuned Resnet-18 model. The fourth column represents the binary masks of the selected tiles by applying WEEP to the tiles ranked by attention scores from the TransMIL model using UNI as the feature extractor. The fifth column represents the binary masks of the selected tiles by applying WEEP to the tiles ranked by attention scores from the atten-MIL model using UNI as the feature extractor.

# Discussion

In this study, we proposed WEEP, a method that provides a spatial tile-level interpretation of deep learning models for prediction modelling of histopathology whole slide images. WEEP is directly linked to the WSI level prediction by a deep learning model and is based on the ranking of the tile-level predictions together with the application of a backward selection strategy to define the subset of tiles driving the assignment of a classification label.

WEEP was demonstrated on CNN and ViT-based models to classify histological grade 1 vs 3 in invasive breast cancer patients and evaluated on three different tile-to-slide aggregation functions i.e. 75th percentile of the tile level prediction scores, the trainable attention-based MIL (atten-MIL) model and transformer-based MIL model (TransMIL). We demonstrated the methodology and visually explored the selected regions in the histological grade 3 patients.

CAM-based approaches are common for pixel-level image interpretation but lack a direct connection to the models' assignment of predictions and decision-making. While the pixel-level information can be of importance to subjectively interpret the model and image patterns learned, areas that are directly linked to how classification labels are assigned are not provided by CAM-based methodologies. Moreover, different CAM-based approaches provide different saliency maps for the same CNN model, which raises questions about the reliability of these methods in offering robust interpretation, especially when semantic pixel-level annotations are not available by definition (e.g. as is the case for prognostic models or other patient outcome-oriented tasks) so that the methods cannot be validated by ground truth annotations (21). In our application domain of histopathology images modelled as tiles, there is also a direct limitation as CAM-based methods only offer interpretability on individual tiles, making it even harder to identify areas (tiles) across multiple instances in a single WSI that are relevant for assignment of the class label (22).

MIL-based methods utilising the tile-level class probabilities to aggregate the WSI-level prediction score are hard to interpret since it is unclear to establish the thresholding on tile-level class probabilities to identify the discriminatory tiles in the WSI. Occlusion Sensitivity Analysis (OSA) have been reported to find discriminatory regions locally (in the tiles) by applying the moving mask in the different regions of the tiles and observing the model's sensitivity (tile-level prediction score) to the different masked regions in a tile (23). It provides an approach for the local interpretability of the selected tiles if needed but does not have a direct association with the WSI-level classification label. Secondary trainable models for tile-to-slide aggregation like the attention-based approach by Lu et al. provide attention weights to directly visualise the highly discriminatory tiles involved in the classification of the WSI (15). However, precise identification of the regions that are needed for the slide-level classification is not possible.

In the specific classification scenario of breast cancer histological grading assessment that includes high inter-observer (24) and inter-lab variability (25), pathologists assign the NHG label on the WSI and it is not possible to obtain well-defined pixel/tile level annotation by definition. To address this problem, multiple MIL-based weakly-supervised CNN-based approaches have been developed especially for the histological grade 1 vs 3 binary classification from the H&E WSIs (11–14). However, it is important to understand the underlying regions recognised by the models and here, the WEEP methodology can be potentially beneficial in providing regions of interest that are directly associated with the classification of the WSI. It is important to note that the WEEP identifies the minimal areas of the WSI that are needed for a particular classification conditional upon the model. If the WEEP selected area is excluded, the model would provide a different classification result. It should not be interpreted as if this area is the only potentially relevant area, but it is a criterion to identify the areas that drive a particular classification decision conditional upon a specific deep learning model.

WEEP has strengths and limitations. One of the main limitations is that the interpretation is provided on the tile level, whereas CAM-based methods are focused on pixel-level interpretations. Secondly, WEEP utilises the underlying model-based ranking of the tiles, which is a potential constraint as the results are conditional upon the model. The major strength lies in the simplicity and empirical approach to defining the direct association of the tiles to the predicted WSI label. The method is also agnostic to the feature extractor and the tile-to-slide aggregation function if the ranking of the tiles associated with the class label can be retrieved. The method can also be extended to the multiclass and regression problems. In the multiclass scenario, WEEP can be applied to the tile-level class probabilities of the predicted class the WSI. In the regression objectives, it is possible to apply WEEP until the WSI prediction intersects with the e.g. mean (or median) of the predicted response variable.

## Conclusions

The proposed WEEP method provides a direct selection of tiles and regions that could be utilised to visually interpret the decision-making of such tile-based CNN classification models. The selected regions can be studied further to understand different tissue morphologies used by the model for classification and could be used to determine novel morphological patterns or confirm the learning of the existing associated morphologies by the CNN-based model.

# List of abbreviations

**WSI:** Whole Slide Image

**CNN:** Convolutional Neural Network

**CAM:** Class Activation Maps

**Grad-CAM:** Gradient-weighted class activation maps

**WEEP:** Wsi rEgion sElection aPproach

**MIL:** Multiple Instance Learning

**H&E:** Hematoxylin and Eosin

**CV:** Cross-Validation

**OSA:** Occlusion Sensitivity Analysis

# Declarations

## Ethics approval and consent to participate

The study has approval by the regional ethics review board (Stockholm, Sweden)

## Consent for publication

Not Applicable

## Availability of data and materials:

Data in the study cannot be made publicly available due to legal constraints. Reasonable access requests to the corresponding author will be considered. The implementation of the WEEP methodology is available at: https://github.com/rantalainenGroup/WEEP.

**Competing interests:**

MR is a shareholder of Stratipath AB. All other authors have declared no conflicts of interest.

**Funding:**

This work was supported by funding from the Swedish Research Council, Swedish Cancer Society, Karolinska Institutet, ERA PerMed (ERAPERMED2019-224-ABCAP), VINNOVA and SweLIFE (SwAIPP project), MedTechLabs, Swedish e-science Research Centre (SeRC) - eCPC, Stockholm Region, Stockholm Cancer Society and Swedish Breast Cancer Association.

**Authors' contributions:**

AS: Data curation, Formal Analysis, Visualisation, Preparing draft of manuscript, Editing manuscript

BL: Supervision, Preparing draft of manuscript, Editing and approval of manuscript

MR: Conceptualisation, Methodology, Visualisation, Funding acquisition, Supervision, Preparing draft of the manuscript, Editing and approval of manuscript

**Acknowledgements:**

Not applicable